\documentclass[aps,pra,groupaddress]{revtex4}

\usepackage{amsmath,amssymb,graphicx}
\usepackage{framed}
\usepackage{color,times}
\usepackage{xcolor}
\usepackage{ulem}
\usepackage{afterpage}

\definecolor{darkblue}{rgb}{0, 0, 0.8}

\usepackage[colorlinks=true, breaklinks=true, linkcolor=darkblue, citecolor=darkblue, urlcolor=darkblue]{hyperref}
\newcommand{\doilink}[2]{\href{http://dx.doi.org/#1}{#2}}
\newcommand{\beq}{\begin{equation}}
\newcommand{\eeq}{\end{equation}}

\newcommand{\ket}[1]{\left| #1\right\rangle}

\begin{document}

\title{Many-Body Physics with Individually-Controlled Rydberg Atoms}

\author{Antoine Browaeys \& Thierry Lahaye}
\affiliation{Laboratoire Charles Fabry, Institut d'Optique Graduate School, CNRS,
Universit\'e Paris-Saclay, F-91127 Palaiseau Cedex, France}

\date{\today}

\begin{abstract}
Over the last decade, systems of individually-controlled neutral atoms, interacting with each other 
when excited to Rydberg states, have emerged as a promising platform for 
quantum simulation of many-body  problems, in particular spin systems. 
Here, we review the techniques underlying quantum gas microscopes and arrays of optical 
tweezers used in these experiments, explain how the different types of interactions between Rydberg 
atoms allow a natural mapping onto various quantum spin models, and describe recent results that 
were obtained with this platform to study quantum many-body physics.  
\end{abstract}

\maketitle

\section{Many-body physics with synthetic matter}

Many-body physics is the field that studies the behavior
of ensembles of interacting quantum particles. This is a broad area encompassing almost 
all condensed matter physics, but also nuclear and high-energy physics. Despite the immense successes 
obtained over the last decades, many phenomena observed 
experimentally still do not have a fully satisfactory explanation.  
At the origin of the difficulty lies 
the exponential scaling of the size of the Hilbert space with the number of interacting particles.
In practice, the best known \textit{ab-initio} methods allow calculating the evolution 
of less than 50 particles. 
To investigate relevant questions involving a much larger number of particles 
(after all even 1~mg of usual matter contains already $10^{18}$ atoms!), 
one must rely on approximations, and the art of solving the many-body problem 
largely relies on mastering them.
However, using approximations is not always possible and  
it may be hard to assess their range of validity. 
One approach to move forward was suggested by Richard Feynman~\cite{Feynman82} 
and consists  in building a synthetic quantum system in the lab, implementing a model of 
interest for which no other way to solve it is known. 
The model may be an approximate description of a 
real material, but it can also be a purely  abstract one. 
In this case, its implementation leads to the construction of an 
artificial many-body system, which becomes an object of study in its own. 
One appealing feature of this approach is the ability to  
vary the parameters of the model in ranges inaccessible
otherwise, thus providing a way to better understand their respective influence. 
For example, if one is interested in the influence of interatomic interactions on
the phase of a given system, synthetic systems become interesting as they allow varying their
strength in a way which is usually impossible in real materials. 
The approach introduced by Feynman is usually referred to as quantum simulation \cite{Lloyd1996,RMP_Qsim}, 
but it can be viewed more generally as exploring many-body physics with synthetic systems: in the  same way 
chemists design new materials exhibiting interesting properties (such as magnetism, superconductivity\ldots),
physicists assemble artificial systems and study their properties, 
with the hope to observe new phenomena. 

For a long time, this idea remained  theoretical as the experimental control 
over quantum objects was not advanced enough. The situation changed radically in the last 20 years
with the development of experimental techniques allowing to control the quantum state of individual 
quantum objects, be they atoms, molecules, ions, photons, or even artificial atoms such as quantum dots, 
superconducting circuits or excitons in semi-conductors, to name a few~\cite{RMP_Qsim}. For all 
these platforms, physicists designed sets of tools allowing the control 
of individual ``atoms'', as well as the ability to tune their interactions. 
This led to the idea of programmable quantum simulation
where all the parameters of the Hamiltonian one wants to implement are tunable.  
But this synthetic systems can also be viewed as machines able to prepare quantum states 
useful for many applications. For example, they can generate large entangled states, whose correlations
are useful to beat the standard quantum limit, hence leading to clocks or sensors with enhanced precision
\cite{Giovannetti2011}.  
In the long-term, they could lead to quantum computers, 
with each ``atom'' carrying a quantum bit \cite{Nielsen,Preskill2012}. 
Interestingly, machines able to implement spin models could be useful to answer  
computationally hard problems well beyond physics, such as combinatorial optimization problems 
(one prominent example being the traveling salesman's problem). Many of these optimization problems can 
be recast as Ising models~\cite{Lucas2014}, that most quantum simulators implement naturally. By varying the 
parameters on the experiment, one could drive the system into a state encoding the solution 
of the problem. 

Among all the platforms being developed (many of them have been  reviewed recently
\cite{Review_QS_ion,Review_QS_atom,Review_QS_Supercond,Review_QS_photon,Review_Interacting_photon}), 
this article will focus on ensembles of individual 
atoms trapped in optical lattices or in arrays of microscopic 
dipole traps separated by a few micrometers.
In this platform, the atoms are almost fully controllable
by optical addressing techniques. To make them interact at distances larger than a micrometer, 
they are excited to Rydberg states, i.e. states with large 
principal quantum numbers $n$ \cite{Gallagher1994,Sibalic2018}. 
When in this state, they feature two important properties. 
First, their lifetime, scaling as $n^3$, is much longer 
than for low lying transitions (typically in the 100\,$\mu$s range for $n\approx 50$). 
Second, they exhibit large dipole moments between states $n$ and $n-1$ 
with opposite parity, scaling as $n^2$. 
This leads to large interaction strengths $V$, corresponding to frequencies $V/h\gtrsim 1$\,MHz 
for $n\approx 50$ at distances around 5\,$\mu$m. 
As we will see in this review, these ensembles of interacting atoms naturally implement
spin models, one of the simplest (and probably most thoroughly studied) many-body systems.

After introducing the concept of Rydberg blockade and the techniques used to prepare and manipulate arrays 
of single atoms, we describe the various types of interactions at play between Rydberg atoms. 
We then review quantum simulation experiments dealing with the Ising and XY spin models, 
and conclude by discussing the perspectives opened by the recent developments of the field. 

\section{Rydberg blockade}

The study of Rydberg atoms played an important role in the early days of atomic physics 
and in the development of quantum mechanics. 
A second `golden age' of Rydberg physics started when tunable lasers became available 
in the 1970's, where their strong coupling to electromagnetic fields made Rydberg atoms an 
ideal test-bed for understanding atom-light interactions, culminating with the birth of cavity 
quantum electrodynamics~\cite{Haroche2013}. 
At this stage,  interactions between Rydberg atoms, although observed as early as 
1981 in dense Rydberg gases~\cite{Raimond1981}, did not play a crucial role. 
This changed at the end of the 1990's, when progress in laser cooling of atoms allowed 
for the realization of \textit{frozen} Rydberg gases~\cite{Anderson1998,Mourachko1998}, 
in which thermal motion is negligible over the timescales where interactions take place. 
Soon after, it was proposed that the  strong interactions between 
Rydberg atoms could be harnessed to implement fast and robust quantum gates between 
neutral atoms \cite{Jaksch2000,Lukin2001}. 
The key ingredient for this implementation is the so-called Rydberg blockade
(see Box 1), where the  interaction prevents the simultaneous Rydberg excitation of two nearby atoms. 
This allows for conditional logic, as the excitation of a second atom is governed by the excitation 
of a first one \cite{Saffman2010}.  

However, at the time, the control of neutral atoms at the individual level was still in its 
infancy~\cite{Schlosser2001}, 
and only a few groups took up the challenge to demonstrate the Rydberg blockade  
between individually controlled atoms. 
This was finally achieved in 2009~\cite{Urban2009,Gaetan2009,Wilk2010,Isenhower2010}. 
In the meantime, many groups had observed clear effects of the Rydberg blockade 
in large ensembles of atoms without individual control (see \cite{Comparat2010} and references therein). 
One soon realized that the theoretical description of these systems naturally mapped onto that 
of the quantum Ising model \cite{Robicheaux2005,Weimer2008,Weimer2010,Olmos2009,Lesanovsky2011}, 
one of the simplest models used to describe quantum magnetism. 
This suggested that systems of neutral atoms in the Rydberg blockade regime could 
be used for quantum simulation, provided individual control of large number of atoms would be available.  

In parallel, the progress in the manipulation and 
detection of individual neutral atoms has made tremendous progress, 
either by using quantum gas microscopes \cite{Bakr2010,Sherson2010} 
or by creating arrays of optical tweezers \cite{Nogrette2014,Lee2016,Barredo2016,Endres2016}. 
Combined with Rydberg excitation to induce controllable interactions between the atoms, 
this provides an almost ideal platform to realize quantum spin models, as we will see below. 
 
\definecolor{shadecolor}{rgb}{0.8,0.8,0.8}
\begin{shaded}
\noindent{\bf Box 1 $|$ The Rydberg blockade}
\end{shaded}
\vspace{-9mm}
\definecolor{shadecolor}{rgb}{0.9,0.9,0.9}
\begin{shaded}
The strong interactions between atoms excited to Rydberg state can be 
exploited to suppress the simultaneous excitation of two atoms and to generate entangled states, in a regime 
called \textit{Rydberg blockade}. Consider a resonant laser field coherently 
coupling the ground state $\ket{g}$ and a given 
Rydberg state $\ket{r}$, with a Rabi frequency $\Omega$ (Figure B1.a). 
In the case of two atoms separated by a distance $R$ (Figure B1.b), 
the doubly excited state $\ket{rr}$ is shifted in energy by the quantity $C_6/R^6$ 
due to the van der Waals interaction (all the other pair states have an energy nearly independent of $R$). 
We assume that the blockade condition $\hbar\Omega\ll C_6/R^6$ 
is fulfilled, i.e. $R\ll R_{\rm b}$ where the \textit{blockade radius} is defined by 
$R_b=\left(C_6/\hbar \Omega\right)^{1/6}$. Then, starting from the ground state $\ket{gg}$, 
the system evolves to the collective state $\ket{\psi_+}=\left(\ket{gr}+\ket{rg}\right)/\sqrt{2}$ 
with a coupling $\sqrt{2}\Omega$. 
The coupling to $\ket{rr}$ is now non-resonant and thus suppressed. 
This leads to a collective Rabi oscillation at the frequency $\sqrt{2}\Omega$ 
between $\ket{gg}$ and the entangled state $\ket{\psi_+}$.

The above considerations can be extended to an ensemble of $N$ atoms 
all included within a blockade volume. In this case, at most one Rydberg excitation 
is possible, leading to collective Rabi oscillations with an enhanced frequency $\sqrt{N}\Omega$ 
between the collective ground state $\ket{g\ldots g}$ and the entangled state 
$\sum_i\ket{g\ldots gr_ig\ldots g}/\sqrt{N}$ where the Rydberg excitation is shared among all the atoms. 
In the case of a system whose size is larger than the blockade radius (Figure B1.c), 
several Rydberg atoms can be excited, but their positions will be strongly correlated due to the blockade constraint, 
giving rise to a complex many-body dynamics.
\begin{center}
\includegraphics[width=10cm]{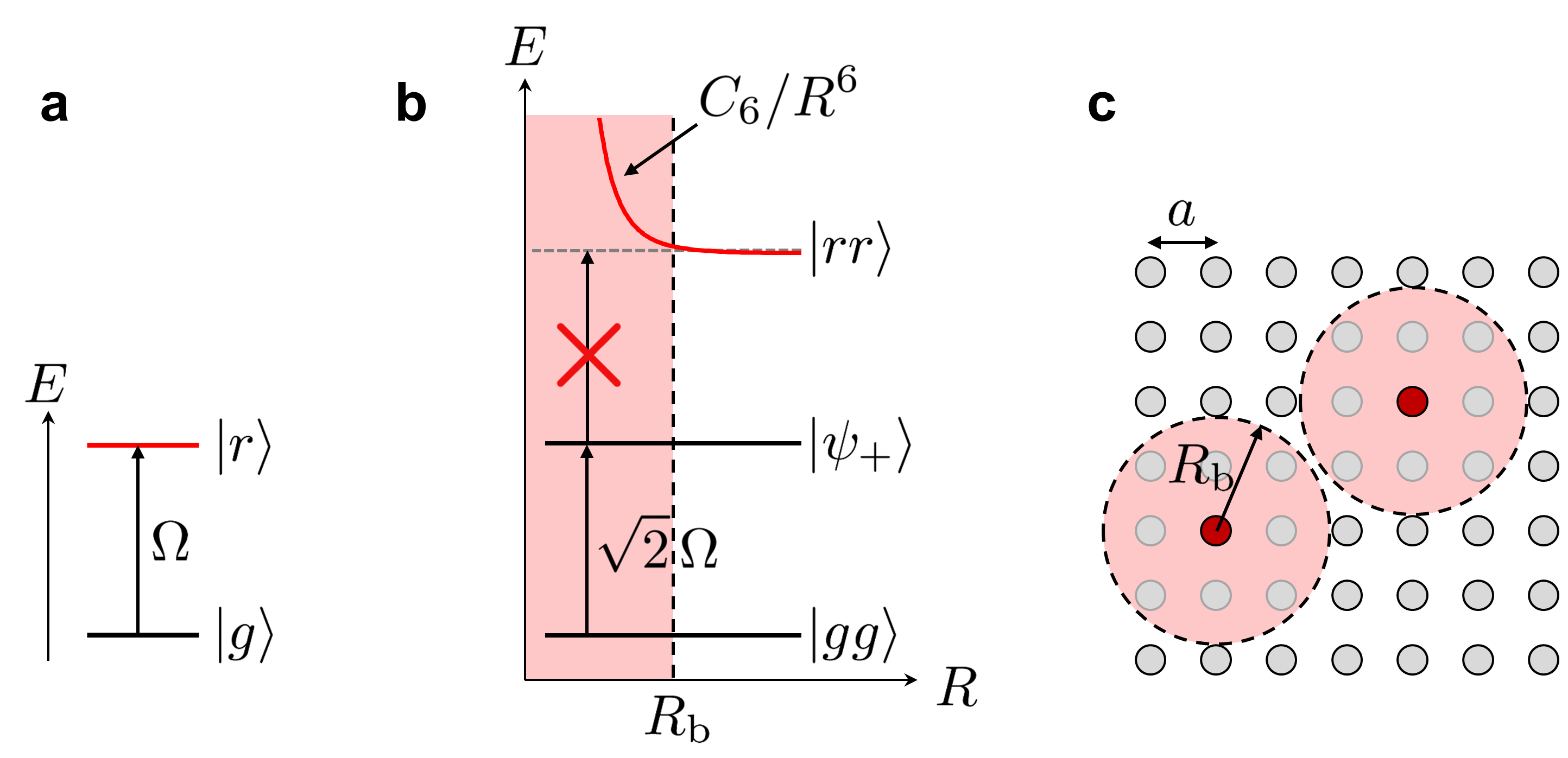}
\end{center}
\small {\noindent{\bf Figure B1 $|$ The Rydberg blockade.} 
{\bf a:} The ground and Rydberg states $\ket{g}$ and $\ket{r}$ are coupled by 
a resonant laser with Rabi frequency $\Omega$. 
{\bf b:} For two atoms separated by a distance $R<R_{\rm b}$, 
the collective ground state $\ket{gg}$ 
is coupled only to $\ket{\psi_+}=\left(\ket{gr}+\ket{rg}\right)/\sqrt{2}$, 
but not to $\ket{rr}$, which is shifted out of resonance by the van der Waals interaction. 
{\bf c:} In a large ensemble of atoms, e.g. a regular array with spacing  $a$, 
an atom excited in $\ket{r}$ (red dot) prevents the excitation of all the atoms 
contained in a sphere of radius $R_{\rm b}$. }
\end{shaded}

\section{Arrays of individual atoms}

\begin{figure*}
\includegraphics[width=14cm]{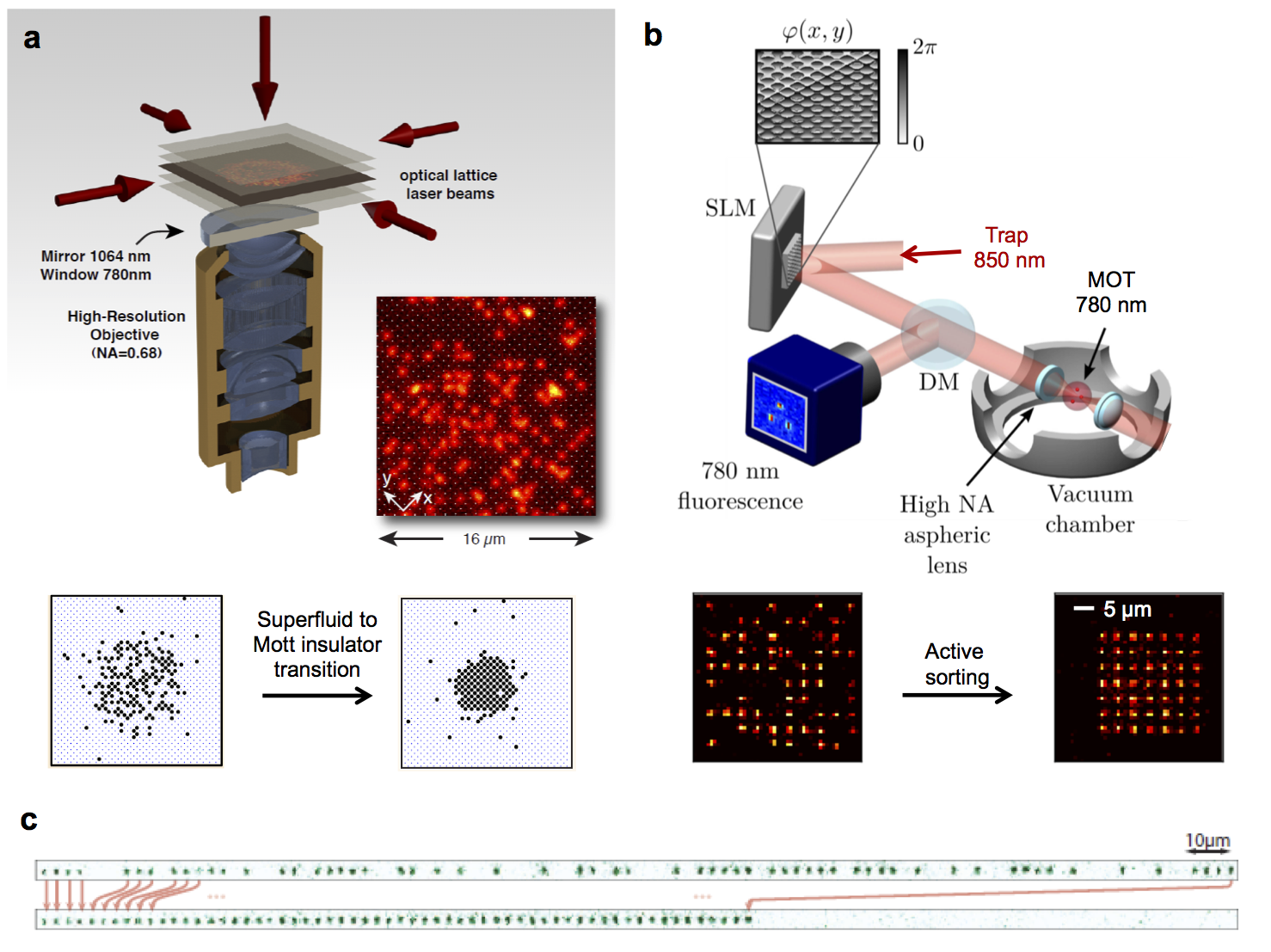}
\caption{{\bf Experimental platforms for realizing arrays of individually controlled neutral atoms.} 
{\bf a:} In a quantum gas microscope, a high-numerical-aperture objective is used to observe 
the fluorescence of ultracold atoms trapped in an optical lattice obtained by interfering 
several laser beams. To achieve a filling of exactly one atom  per site, one drives the superfluid 
to Mott-insulator transition~\cite{Sherson2010}. 
{\bf b:} In the tweezers array platform, a spatial-light modulator (SLM) imprints an 
appropriate phase on a trapping beam prior to focusing with a high-numerical-aperture lens, 
resulting in arrays of traps with almost arbitrary geometries~\cite{Nogrette2014}. 
Single, laser-cooled atoms are loaded in the optical tweezers from a magneto-optical trap, 
resulting in a random loading array at half filling, which can be actively reordered into a 
target array using a moving optical tweezers~\cite{Barredo2016}. Bottom: single shot fluorescence image 
of an array of traps before and after assembly. 
{\bf c:} Alternatively, in one dimension, the tweezers can be generated using 
an acousto-optic deflector fed with multiple radio-frequency tones, 
which allows rearranging the atoms in a single step~\cite{Endres2016}.}
\label{fig:fig1} 
\end{figure*}

The first experimental platform that has allowed the control of ordered assemblies of 
neutral atoms at the single-particle level became available in 2010 and is the 
\textit{quantum gas microscope}~\cite{Gross2017} (Figure~\ref{fig:fig1}a). 
This ``top-down'' approach relies on the loading of a 
two-dimensional ultracold atom cloud, typically a Bose-Einstein condensate 
---although fermions can also be used--- 
into an optical lattice, \textit{i.e.} the periodic optical potential obtained by interfering several laser beams. 
Atoms can tunnel between neighboring sites of the lattice, and when the on-site contact 
interaction between the atoms overcomes the kinetic energy given by the tunneling rate, 
the system undergoes a superfluid to Mott insulator transition~\cite{Greiner2002}. 
Deep in the Mott phase, the system is characterized by a fixed number of atoms per site, 
which can be exactly one for an appropriate choice of parameters. 
In order to obtain single-site resolution when imaging the atomic fluorescence, 
a high numerical aperture microscope objective is required, 
as two neighboring sites are separated by typically 
500~nm~\cite{Bakr2010,Sherson2010}. In this way, one realizes two-dimensional 
square arrays of up to a few hundred single atoms, with filling fractions that can exceed 95\%. 
Individual control of the atoms can be achieved by applying local light shifts tailored 
with a spatial light modulator (SLM) such as a Digital Micro-mirror Device~\cite{Weitenberg2011}. 

More recently, a novel, ``bottom-up'' platform has emerged based on arrays of 
optical tweezers (Figure~\ref{fig:fig1}b). Trapping of a single laser-cooled atom in a tightly 
focused dipole trap, or optical tweezers, was demonstrated as early as 2001~\cite{Schlosser2001}. 
An optical dipole trap with a tight focus of about 1~$\mu$m is immersed in a magneto-optical trap (MOT). 
In the course of its random motion, an atom from the MOT cloud enters the optical tweezers and 
gets trapped there. Since it is still under the illumination of the near-resonant MOT beams, 
it continuously scatters fluorescence light, that can be collected on a sensitive camera, 
thus signaling the presence of an atom in the tweezers. If now a second atom enters the trap, 
very fast light-assisted collisions give rise to the almost immediate loss of both atoms, and 
the tweezers is empty again. Therefore the number of atoms in the microtrap is either zero or one: 
this is a \textit{single-atom source}, albeit  non deterministic since one cannot 
predict when the trap is occupied. As the same random event (namely the random entering of 
an atom in the trap) induces the transitions from one to zero atom and from zero to one atom, 
the occupation probability is $p\sim 1/2$. 

The next step is to produce arrays of microtraps. One method relies on a spatial 
light modulator \cite{Bergamini2004,Nogrette2014}, 
which imprints an appropriate phase pattern on the trapping beam before focusing, 
thus allowing the realization of almost arbitrary arrays of traps in the focal plane of the objective.
Other methods use arrays of microlenses~\cite{Dumke2002,Schlosser2012} 
or interference techniques~\cite{Piotrowicz2013}. 
However, for a long time, the stochastic loading of microtraps limited the use of this platform to just a few atoms, 
since the probability of having all $N$ traps simultaneously filled decreases as 
$1/2^N$. Careful engineering of the light-assisted collisions in order to lose just one atom of the 
pair~\cite{Grunzweig2010,Lester2015} was shown to enhance $p$ to values up to $\sim 0.9$, 
but the probability $p^N$ of an $N$-trap array to be defect-free still decreases very fast with $N$. 
This problem was circumvented  simultaneously in 2016 by three groups. 
The idea is to start from a large array with $2N$ traps, load it randomly with $\sim N$ atoms, 
take an image of this configuration, and finally actively sort the atoms into an ordered, target configuration
\cite{Miroshnychenko2007}. 
This was achieved by two methods. In the first one, loaded tweezers are dynamically moved 
using acousto-optic deflectors to assemble one-dimensional chains~\cite{Endres2016} (Figure~\ref{fig:fig1}c) as well as 
two~\cite{Kim2016} and three-dimensional~\cite{Lee2016} arrays  
by slowly varying the phase pattern of the SLM creating the array. 
In the second method, atoms are moved one at a time using a moving optical tweezers 
to catch and release atoms within a fixed two-dimensional array produced by a SLM~\cite{Barredo2016}
or microlenses~\cite{Mello2019}. More recently, the latter technique was extended 
to the assembling of three-dimensional arrays~\cite{Barredo2018}. 
The assembling approach offers a fast repetition rate 
of the experiment (a few per second), filling fractions in excess of 98\% even in 
large arrays, and a great flexibility in geometry; 
the atom number reached so far is around one hundred atoms. 
This sorting of atoms has also been applied in two and three-dimensions in optical lattices with large spacing
between sites~\cite{Nelson2007,Kumar2018}.
 
\section{Interactions between Rydberg atoms and mapping onto spin systems}

The large electric dipole moment of atoms excited to Rydberg states 
leads to strong dipole-dipole interactions between them~\cite{Browaeys2016,Sibalic2017,Weber2017}, 
even for atoms separated by micrometer-large distances as in the lattices and arrays described above. 
Two types of interaction naturally occur between two Rydberg atoms (see Box~2). In the context 
of quantum simulation, this leads to a mapping onto different spin models. 

Let us consider first the case where the two atoms are placed in the {\it same} Rydberg state. 
There, the dipole-dipole interaction leads to the van der Waals interaction, which induces
an energy shift of the pair state $|rr\rangle$ scaling as $C_6/R^6$. This shift occurs only when
the two atoms are both excited to the Rydberg state. If we now map the ground and Rydberg state 
$|g\rangle$ and $|r\rangle$ of each atoms onto a spin $1/2$ following 
$\ket{\downarrow}=\ket{g}$ and $\ket{\uparrow}=\ket{r}$ the Hamiltonian of an ensemble of 
atoms driven by a coherent laser (Rabi frequency $\Omega$, frequency detuning $\delta$)
is \cite{Robicheaux2005}
\begin{equation}
H=
\frac{\hbar\Omega}{2} \sum_i \sigma_x^i
-\hbar\delta \sum_i n_i
+\sum_{i< j}V_{ij} n_i n_j, \ \textrm{with}\ V_{ij}=\frac{C_6}{R_{ij}^6}.
\label{eq:vdw}
\end{equation}
Here $n_i$ is the operator counting the number of Rydberg excitation at site $i$, and
$\sigma_x$ is the usual Pauli matrix. As $n_i$ is related to the $\sigma_z$ Pauli matrix by $n_i=(\sigma_z^i+1)/2$, 
Eq.(\ref{eq:vdw}) has the form of the quantum Ising model, with a transverse field $B_\bot \propto \Omega$, 
a longitudinal field $B_{||}\propto-\delta$ 
and Ising couplings $J_{ij}$ decaying as $1/R_{ij}^6$ with the distance. 
In practice, for the alkali atoms used so far, the laser excitation leading to the Rabi frequency $\Omega$ is 
often achieved on a two-photon
transition, with one of the laser in the infra-red and the other one in the blue, far-detuned 
with respect to an intermediate state in order to avoid spontaneous emission~\cite{Johnson2008,Myroshnychenko2010}.  
In this way, the excitation is coherent to a good approximation. The Hamiltonian of Eq.(\ref{eq:vdw}) assumes that 
the excitation laser covers uniformly the atomic array, but owing to the single-site addressability, the detunings and Rabi
frequency can be made site dependent by adding local laser control~\cite{Labuhn2014}.  
Finally, it is also possible to realize the quantum Ising model by using a technique called Rydberg dressing and 
encoding the two spin states in ground-state long-lived levels~\cite{Jau2016}. 
In this case, the couplings $J_{ij}$ have  a soft-core spatial dependence (see Box~2).

We now consider the second case where the atoms are prepared in 
two {\it different} Rydberg states that are dipole-coupled, 
such as $|nS\rangle$ and $|nP\rangle$,  separated by a 
transition frequency typically in the 10~GHz range. There, 
the dipole-dipole interaction gives rise to a coherent exchange 
of the internal states of the atoms and the interaction potential 
scales like $C_3/R^3$. The mapping onto a spin-$1/2$ model
is then $\ket{\downarrow}=\ket{nS}$ and $\ket{\uparrow}=\ket{nP}$. 
Microwave radiation can be used to 
manipulate the spin and thus acts as an external magnetic field. 
The Hamiltonian for a system of atoms then reads:
\begin{equation}
H=
\frac{\hbar\Omega_{\mu \rm w}}{2} \sum_i \sigma_x^i
-\frac{\hbar\delta_{\mu \rm w}}{2} \sum_i \sigma_z^i
+\sum_{i\neq j}\frac{C_3}{R_{ij}^3}\left(\sigma_+^i\sigma_-^j+\sigma_-^i\sigma_+^j\right),
\label{eq:xy}
\end{equation}
which is the XY spin Hamiltonian with transverse and longitudinal fields given 
by the Rabi frequency $\Omega_{\mu \rm w}$ and the detuning $\delta_{\mu \rm w}$
of the microwave field. 

Both the Ising and the XY Hamiltonians have been extensively studied 
in the last 60 years in various contexts, such as magnetism and excitation transport. 
However many important open questions remain 
the subject of active research, such as the nature of the phase diagram when 
the spins are placed in arrays featuring geometrical frustration, the dynamics of the 
system after a quench ({\it}i.e. the sudden variation of one parameter of the Hamiltonian), the role 
of disorder in the couplings, their combination with situations where topology plays a role\ldots
Besides, as explained in the introduction, many combinatorial optimization problems can be mapped onto spin models
\cite{Lucas2014}, and their interest is therefore beyond the traditional realm of many-body physics. 
All these questions can be be studied using the Rydberg platforms described here, as we now show.

\definecolor{shadecolor}{rgb}{0.8,0.8,0.8}
\begin{shaded}
\noindent{\bf Box 2 $|$ Interactions between Rydberg atoms}
\end{shaded}
\vspace{-8mm}
\definecolor{shadecolor}{rgb}{0.9,0.9,0.9}
\begin{shaded}
Two atoms separated by a distance $R$ much larger than the size of the electronic 
wavefunction interact mainly \textit{via} the electric dipole-dipole Hamiltonian
$\hat{V}_{\rm dd}\sim \hat{d_1}\hat{d_2}/ (4\pi\varepsilon_0 R^3)$
where $\hat{d}_i$ is the electric dipole moment of atom $i$. 
The effect of this Hamiltonian on a pair of Rydberg atoms depends on how the pair is prepared. 

In the most common case,  the two atoms are excited in the same Rydberg level, 
for instance $\ket{nS}$. In this case, $\hat{V}_{\rm dd}$ 
has no effect at first order in perturbation theory, 
as an atomic state has a vanishing average electric dipole moment. 
The effect of the interaction is thus of second order: 
the pair state $\ket{nS,nS}$ is coupled via $\hat{V}_{\rm dd}$ to other pair states of opposite parity 
differing in energy from $\ket{nS,nS}$ by a quantity $\Delta$ (Figure~B2a) 
with a matrix element $V\propto d_1d_2/R^3$. 
This gives rise to a van der Waals shift of the considered pair state scaling as 
$V^2/\Delta\propto C_6/R^6$~\cite{Reinhard2007}. The $C_6$-coefficient 
is on the order of $ d^4/\Delta$, and
thus scales roughly as $n^{11}$ since $d\sim n^2$ and $\Delta\sim n^{-3}$. 
The dependence on distance of the van der Waals interaction was directly measured for the cleanest 
system of a pair of single atoms at controlled positions in \cite{Beguin2013}, 
as was its angular dependence in~\cite{Barredo2014}. 

The van der Waals interaction between two Rydberg atoms is huge: 
it can reach tens of MHz for atomic separations of several micrometers. 
However Rydberg states have a lifetime on the order of a few hundreds of microseconds. 
In order to reach much longer lifetimes, at the expense of reducing the interaction strength, 
the idea of \textit{Rydberg dressing} has been proposed 
\cite{Bouchoule2002,Pupillo2010,Johnson2010,Balewski2014,Glaetze2014}. 
It consists in driving off-resonantly the transition from the ground to the Rydberg state, 
in a regime where the Rabi frequency is smaller than the detuning. 
The Rydberg state population remains negligible, but the ground state being weakly admixed with 
the Rydberg state, a pair of ground-state atoms acquire sizable interactions, 
with a long-distance tail decaying as $1/R^6$ beyond the Rydberg blockade radius, 
and a flat-top interaction at shorter distances. Figure B2b shows this \textit{soft-core potential}, 
measured on a pair of single atoms trapped in optical tweezers, 
for two different detunings of the dressing laser~\cite{Jau2016}.

On the contrary, when the two atoms are prepared in two different, dipole-coupled Rydberg states, 
such as $\ket{nS}$ and $\ket{nP}$, 
the pair state $\ket{nS,nP}$ is directly coupled to the same-energy state 
$\ket{nP,nS}$ by $\hat{V}_{\rm dd}$ (Figure~B2c). 
This gives rise to new eigenstates $\ket{nS,nP}\pm\ket{nP,nS}$ 
with energies $\pm C_3/R^3$. A pair of atoms initially prepared in $\ket{nS,nP}$ will 
coherently evolve into $\ket{nP,nS}$ and back, with a ``flip-flop'' oscillation 
frequency $\propto R^{-3}$ (Figure~B2.c, adapted from \cite{Barredo2015}). 
Moreover the interaction is anisotropic, varying as $1-3\cos^2\theta$ with $\theta$ 
the angle between the internuclear axis and the quantization axis (inset of figure~B2.c).

Finally, online calculators are now available to compute the interactions in all regimes, 
including in the presence of external electric and magnetic fields~\cite{Sibalic2017,Weber2017}.

\begin{center}
\includegraphics[width=11cm]{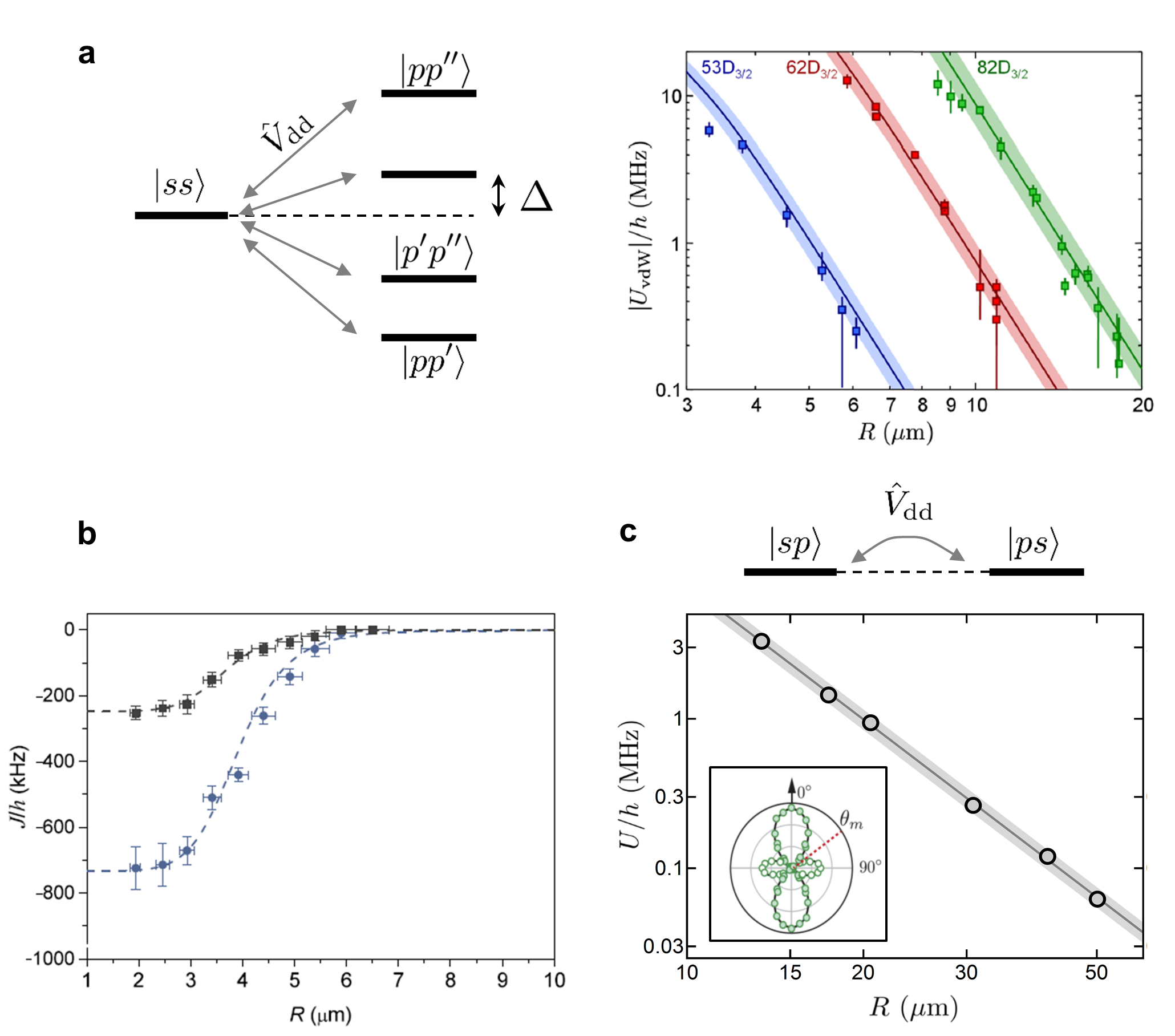}
\end{center}
\small{\noindent{\bf Figure B2 $|$ Interactions between Rydberg atoms.} 
{\bf a:} The van der Waals interaction between two identical Rydberg states, 
for instance $\ket{nS}$ states, arises due to non-resonant dipolar interactions 
with other dipole-coupled Rydberg states. 
The $C_6$ coefficient varies extremely fast with the principal quantum number $n$, as illustrated
on this measurement for two atoms separated by $R$, excited to $nD$ states~\cite{Beguin2013}. 
{\bf b:} Soft-core potential between Rydberg dressed atoms, measured for a pair 
of single atoms, as a function of the atomic separation~\cite{Jau2016}. 
{\bf c:} The resonant dipole-dipole interaction arises when two atoms are in different, 
dipole-coupled Rydberg states, such as $\ket{nS}$ and $\ket{n'P}$. 
It varies as $C_3/R^3$ with the distance \cite{Barredo2015}, with the angular dependence 
$C_3\propto1-3\cos^2\theta$ typical of the dipole-dipole interaction 
(inset, adapted from \cite{deLeseleuc2019}).} 

\end{shaded}

\section{Quantum simulation of the Ising model}

To study many-body systems experimentally one can, for example, vary suddenly one parameter of 
the Hamiltonian and study the resulting dynamics of the closed many-body system.
One can also prepare the ground state using  an adiabatic variation of the parameters of
the Hamiltonian, and study its properties. The experiments performed  on Rydberg quantum simulators in the 
Ising model regime used these two approaches. 

\begin{figure*}
\includegraphics[width=16cm]{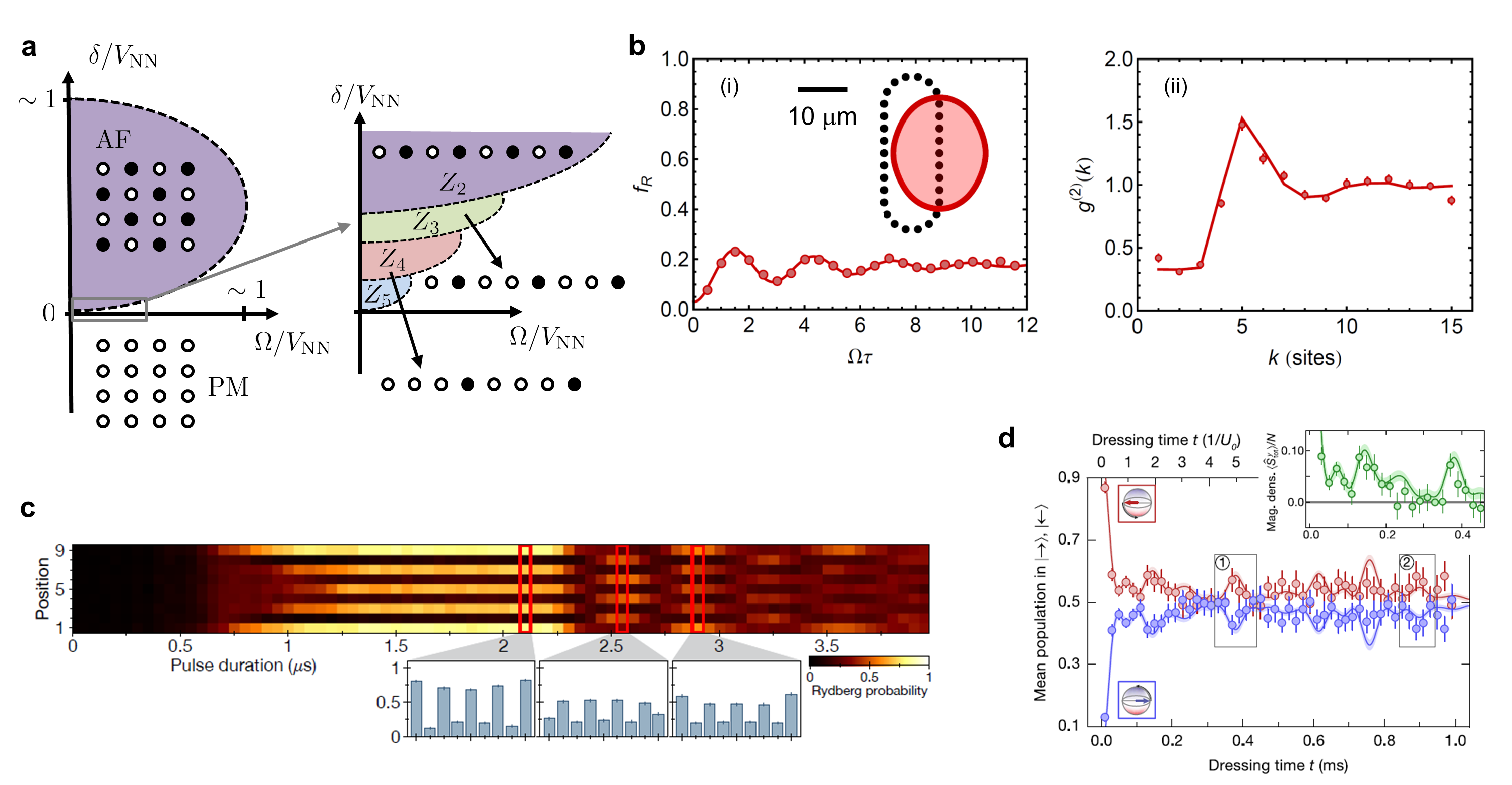}
\caption{{\bf Quantum quench experiments for the Ising model.} 
{\bf a:} Schematic phase diagram of the quantum Ising model for a 2d square or a 1d chain of atoms interacting via the 
van der Waals interaction, showing the paramagnetic (PM) and antiferromagnetic (AF) phases. 
Here $V_{\rm NN}=C_6/a^6$, with $a$ the spacing between atoms (we take $C_6>0$). 
The right side is a zoom of the phase diagram around the critical point $(\Omega=0,\delta=0)$, for the 1d chain.
{\bf b:} Quantum quench in a one-dimensional Ising magnet with periodic boundary 
conditions, starting with all atoms in $\ket{\downarrow}$, for $R_{\rm b}/a\simeq 4$ \cite{Labuhn2016}. 
The shaded area represents the blockade volume. 
The magnetization of the chain 
(i) and the spin-spin correlation function 
(ii) show clear effects of the Rydberg blockade. 
The solid lines are simulations of the Sch\"odinger equation 
without any adjustable parameter. {\bf c:} In \cite{Bernien2017}, 
a quasi-adiabatic sweep was used, in the regime $R_{\rm b}/a\simeq 1.5$, 
to prepare an `antiferromagnet' $\ket{\uparrow\downarrow\uparrow\downarrow\cdots}$. 
This state was then suddenly quenched (at 2.2~$\mu$s) by driving it on resonance, 
resulting in surprisingly long-lived collective oscillations between 
$\ket{\uparrow\downarrow\uparrow\downarrow\cdots}$ and 
$\ket{\downarrow\uparrow\downarrow\uparrow\cdots}$. 
{\bf d:} Rydberg dressing can also be used to study quenches of Ising magnets. 
In~\cite{Zeiher2017}, a chain of 10 atoms all initially prepared in 
$\ket{\leftarrow}=(\ket{\uparrow}+\ket{\downarrow})/\sqrt{2}$ was suddenly 
subjected to the Ising Hamiltonian, giving rise to a dynamical evolution showing 
collapse and revivals of the magnetization along $x$.}
\label{fig:fig2} 
\end{figure*}

Let us first discuss qualitatively the generic phase diagram of the 
quantum Ising model described by Eq.(\ref{eq:vdw}), at zero temperature and for spins placed on 
a chain or on two-dimensional square arrays.   
We first consider the case of nearest neighbor couplings only ($V_{i,i+1}=V$ and $0$ otherwise) 
and  assume $V>0$, so that the interactions favor anti-ferromagnetic ordering.
The phase diagram consists of two regions, a paramagnetic and an anti-ferromagnetic one, 
separated by a quantum phase transition, as represented in Fig. \ref{fig:fig2}a. 
Two limiting cases are easy to understand: for $\Omega, \delta \gg V$, the ground state is paramagnetic, {\it i.e.} 
the spins align along the effective magnetic field; for $\Omega=0$, the phases results from the minimization of the energy 
of the classical configuration. 
When we relax the constraint of nearest neighbor couplings  only (as is the case for a van der Waals interaction)
the diagram exhibits several phases around the line separating the para- from the antiferromagnetic
phases. For example, on a chain, if  $V_{i,i+1}, V_{i,i+2}\gg\delta\gg\Omega\gg V_{i,i+3}$, the ground 
state corresponds to one excitation separated by two ground state atoms ($Z_3$ symmetry). 
This situation corresponds to $R_{\rm b}=2a$, with $a$ the spacing between atoms.
Similarly, $V_{i,i+1}, V_{i,i+2},V_{i,i+3}\gg\delta\gg\Omega\gg V_{i,i+4}$ 
leads to a phase with $Z_4$-symmetry, and so on. 
By controlling the detuning $\delta$ and Rabi frequency $\Omega$, 
one can explore the phase diagram of this Ising model. 

The sudden variation of a parameter in the Hamiltonian (also called a quench), 
is the easiest method to implement experimentally. 
In the case of the Rydberg platform, all the experiments realized so far suddenly applied
the Rydberg excitation laser mimicking the transverse magnetic field, usually at resonance ($\delta=0$),
after having prepared the atoms in their ground state 
({\it i.e.} spin $\ket{\downarrow}$). 
They then measure two quantities relevant to the study of spin systems. 
First, the average magnetization, {\it i.e.} 
the average number of atoms excited to the Rydberg states, or equivalently 
in the spin state $\ket{\uparrow}$. 
Second, the spin-spin correlation function, which is the probability to find a 
Rydberg excitation at site $j$ when one is
already present at site $i$. 
The measurement of this correlation function is only made possible owing to the fact that
the quantum gas microscope and the tweezer array platforms 
allow for single site readout of the atomic state.
The group of I. Bloch and C. Gross at the Max-Planck Institute in Garching 
was the first to implement this method using their
newly built quantum gas microscope~\cite{Schauss2012}. 
They operated in a regime where the blockade radius $R_{\rm b}$
was much larger that the inter-site distance $a\sim 500$\,nm. 
The group could observe the effect of the blockade 
as the fact that Rydberg excitations appeared separated typically by a distance $R_{\rm b}$. 
They could also follow
the dynamics of the appearance of the excitation and compare 
it to the theoretical prediction of the Ising model. Later, the same group has also demonstrated
the $\sqrt{N}$ enhanced coupling  for an ensemble of atoms all within a blockade radius 
for up to 200 atoms~\cite{Zeiher2015}. 

The case $R_{\rm b}\gg a$, as explored by the Max Planck group, 
is an extreme situation where the interaction dominates all the energy scales in the problem. 
It is also interesting to explore the case where the interaction energy 
between neighbouring atoms is on the order of the energy scale 
associated with the transverse field, {\it i.e.} operating at $R_{\rm b}\sim a$. 
The tweezer arrays is naturally in this regime
and our group at Institut d'Optique was the first to explore this 
situation~\cite{Labuhn2016}. As an example, we considered
a one-dimensional chain of $\sim 20$ atoms with periodic boundary conditions 
(see Fig.~\ref{fig:fig2}b). 
After suddenly turning on the excitation laser, we observed the dynamics 
and measured the pair correlation functions. We compared
them to the solution of the Schr\"odinger equation for this many-body system, 
including all the experimental imperfections. 
The agreement with the data is very good, and this theory-experiment 
comparison can be considered as a benchmark of
this Rydberg quantum simulator in  a regime where {\it ab-initio} calculations are still possible.   
In particular, the appearance of a steady state regime for the magnetization 
at long time results from the beating of
all the eigen-frequencies of this interacting system. 
The behavior of the pair-correlations function (also observed in 
two-dimensions by the Max Planck group~\cite{Schauss2012}) with its suppression 
at short distance and its oscillatory behavior at larger distance
is reminiscent of the pair correlation function of a liquid of hard rods 
with an effective size of the particles of $R_{\rm b}$.  We repeated this quench experiment
in a $7\times 7$ arrays \cite{Leseleuc2018PRL}, 
and compared the observed dynamics to a theory, which had to be approximate
as the number of particles involved is too large to allow for {\it ab initio} calculations. 
These studies have been refined by the group of J. Ahn in Korea, 
who considered the appearance of a steady state at long evolution times as
an evidence of thermalization of the system~\cite{Kim2018,Lee2019}.
Finally the group of M. Lukin at Harvard also studied an out-of-equilibrium 
situation, but contrarily to the experiments 
described above, they started from the ground state of the many-body system~\cite{Bernien2017} obtained from 
an adiabatic preparation detailed below. 
By suddenly changing the detuning of the laser
so as to cross a quantum phase transition, they observed non trivial dynamics with long-lived collective oscillations 
that they were able to model (see Fig.~\ref{fig:fig2}c). 

\begin{figure*}
\includegraphics[width=16cm]{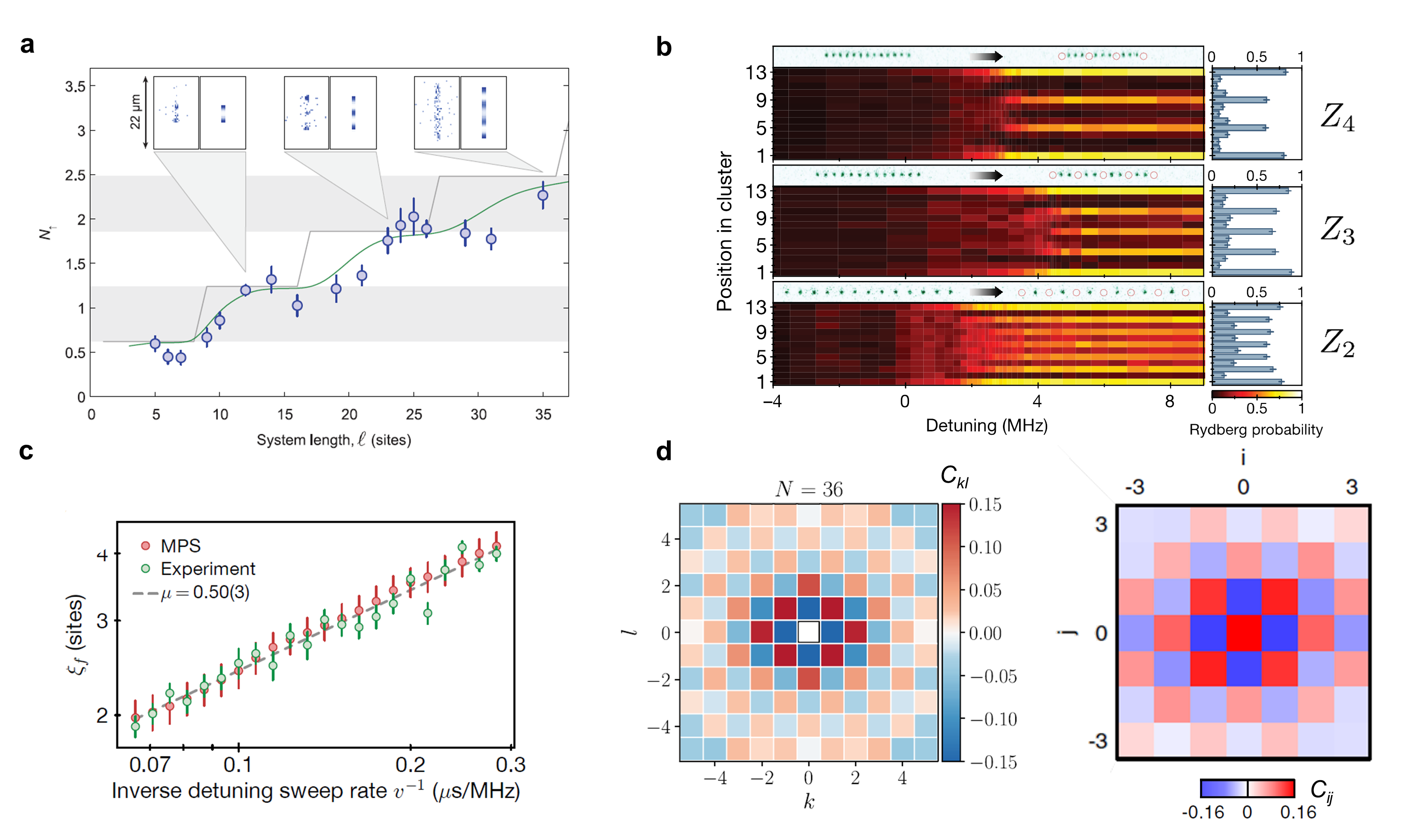}
\caption{{\bf $|$ Quasi-adiabatic sweeps experiments for the Ising model.} 
{\bf a:} ``Crystallization'' of Rydberg excitations in a quasi one-dimensional chain, 
in a regime where the blockade radius is much larger than the lattice spacing $a$. 
When the chain length is increased, the number of excitations 
increases in a step-like manner \cite{Schauss2015}. 
{\bf b}: Adiabatic preparation of the ground state of a 13-atom chain for various values 
of $R_{\rm b}/a$, giving an excitation every second site, every third site, 
or every fourth site, corresponding to different ordered phases Z$_n$ of the system \cite{Bernien2017}. 
{\bf c:} Antiferromagnetic correlation length $\xi$ obtained in a chain of 51 atoms as function of 
the detuning sweep rate~ \cite{Keesling2019}. The slope gives access to 
critical exponents characterizing the quantum phase transition.
{\bf d:} Antiferromagnetic spin-spin correlation function in two dimensional arrays of atoms 
after adiabatic preparation for the experiment of \cite{Lienhard2018} (left) and \cite{GuardadoSanchez2018} (right).}
\label{fig:fig3} 
\end{figure*}

A second method to study many-body systems consists in preparing the ground state of the system. 
In order to do so, one starts from the state where all atoms are prepared in their ground state:
$|\psi_{\rm ini}\rangle=|ggg\ldots\rangle$. By sweeping the Rabi frequency $\Omega(t)$ 
and detuning $\delta(t)$ on a time scale long with respect to the inverse of the energy gap
with the first excited state, the systems is driven adiabatically in the ground state of the 
interacting system for a given final value of 
$\Omega$ and $\delta$. Once again, based on a suggestion by T. Pohl~\cite{Pohl2010}, 
the group of I. Bloch was the first to demonstrate this approach in the regime $R_{\rm b}\gg a$
~\cite{Schauss2015}: they observed a controlled number of excitations separated by $R_{\rm b}$ in
one and two dimensions, a situation reminiscent of a crystal. 
The regime $R_{\rm b}\sim a$, was then explored by three groups. The group of 
M. Lukin in Harvard investigated the one-dimensional case~\cite{Bernien2017} with up to 51 atoms. 
In their case, they varied the ratio $R_{\rm b}/a$ between 2 and 4, so that they could access several 
$Z_n$ phases. The Institut d'Optique group 
\cite{Lienhard2018} and the group of 
W. Bakr in Princeton \cite{GuardadoSanchez2018} explored the two-dimensional case using 
respectively atoms in arrays of tweezers and in optical lattices. Both groups observed the appearance
of antiferromagnetic correlations in their system. The Institut d'Optique group also studied 
the propagation of these correlations during the adiabatic ramp of the parameters.
Let us note that this adiabatic approach becomes harder and harder as the system size increases, as the energy gap 
vanishes at the quantum phase transition. 

Of course, the transition between the quenched and adiabatic regimes is continuous. The group
of Harvard studied what happens at the quantum phase transition when varying the rate a which 
the parameters are varied~\cite{Keesling2019}. 
In doing so they found out that if the ramp is too fast, defects with respect 
to the ideal $Z_n$ symmetry appears. This defects can be studied according to a model introduced
by Kibble and Zurek in the 80's~\cite{Kibble76,Zurek85}. In particular, the correlation length obtained as a function 
of the speed at which the parameters are ramped follows a universal law with a critical exponent 
that can be extracted from the experiment. They found an experimental value in 
excellent agreement with the theoretical prediction from the Ising model. 
Their experiment nicely illustrates that synthetic
quantum systems can be used to measure with high precision the properties of
quantum phase transitions. 

As pointed out at the beginning of this review, Rydberg atoms have lifetimes in the 100's of 
microseconds, a time long enough to observe interaction-driven dynamics. 
However some features such are revivals in the dynamics, occurring at long times for large systems,
may be hard to observe, as
the atoms decay to their ground states before the revival occurs. 
The Rydberg dressing introduced earlier was
proposed as an alternative to circumvent this problem 
\cite{Bouchoule2002,Pupillo2010,Johnson2010,Balewski2014,Glaetze2014}. First experiments in this direction 
were performed in the group of I. Bloch and C. Gross in Garching \cite{Zeiher2016,Zeiher2017}. There, 
the spin is encoded in two hyperfine ground states of Rb atoms. The atoms arranged along 
a chain (containing around 10 atoms) are initially all 
prepared in a superposition state $\propto\ket{\uparrow}+\ket{\downarrow}$. A laser is then switched
on to admix the state $\ket{\uparrow}$ with a Rydberg state $|r\rangle$, thus making the 
atoms interact. The system evolves under the influence of the interactions. After some time, the laser is
switched off and the state of each atom is read out. One can then compute the average magnetization
and the  correlations. The group was able to observe revivals on the magnetization, as shown
in Fig.~\ref{fig:fig2}(d). 
Although this result is very encouraging as it shows the potential of Rydberg dressing,
it remains a bit unclear at the present time how large the system sizes can be, 
as unwanted losses appear during the dressing when the number
of atoms grows~\cite{Goldschmidt2016,Boulier2017,Zeiher2016}.

\section{Quantum simulation using resonant exchange interactions}

\begin{figure*}
\includegraphics[width=16cm]{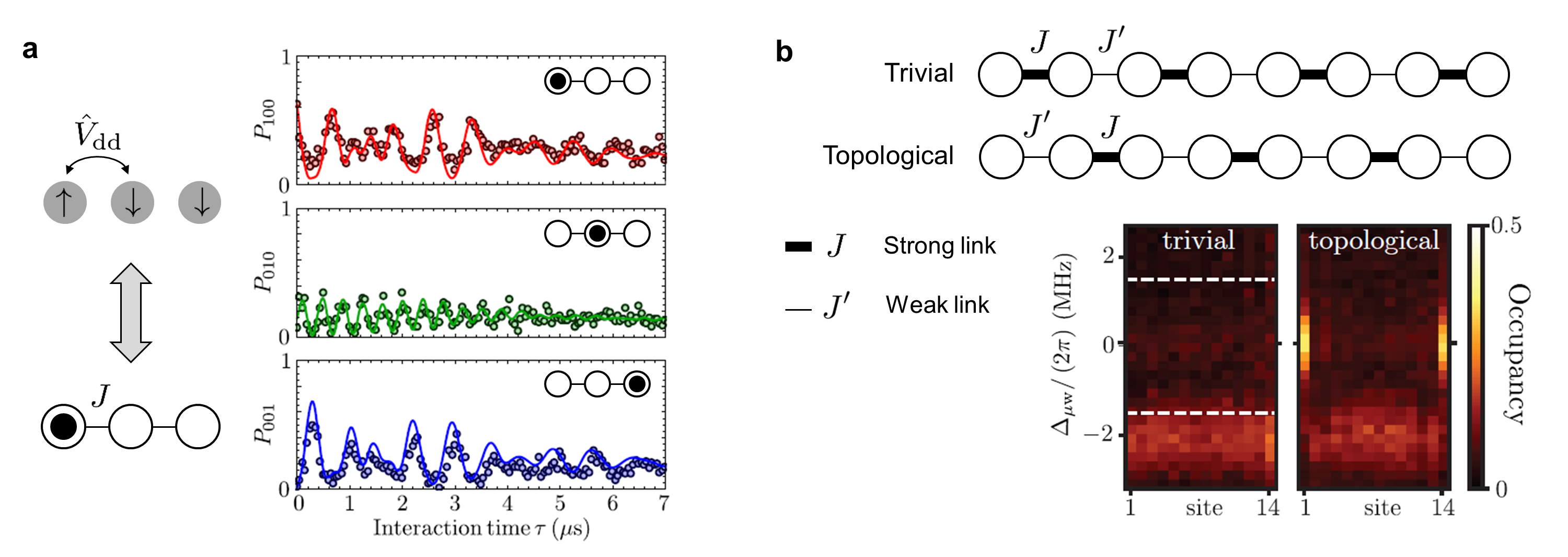}
\caption{{\bf Quantum simulation of the XY model.} 
{\bf a:} Mapping between the XY model and the propagation of a particle from one site to another in a chain.
{\bf b:} Illustration in a chain of 3 atoms, where the initial state is $\ket{PSS}$: evolution of the probability that 
the particle ``$P$'' is on the left, center or right site, as a function of time. One observes quasi revivals. From \cite{Barredo2015}.
{\bf c:} Implementation of the Su-Schrieffer and Heeger model using Rydberg atoms. The model consists of 
a chain with an alternance of strong and weak couplings between neighboring sites. A finite chain of even number of
sites exists in two different configurations (called normal or topological), with the weak or the strong link at the edges. 
{\bf d:} Microwave spectroscopy of the Su-Schrieffer-Heeger model implemented on a chain of Rydberg atoms: 
probability of exciting the atoms from Rydberg state $nS$ to $nP$ 
as a function of the frequency of the microwave, and for different 
positions in the chain. The topological configuration features two at zero energy, corresponding to atoms 
located at the two edges of the chain. From \cite{deLeseleuc2019}.   }
\label{fig:fig4} 
\end{figure*}

As explained above, the resonant dipole-dipole interaction between two Rydberg states naturally realizes
 the XY spin-$1/2$ model. Besides its interest for the study of quantum magnetism, this model is also useful to 
describe transport properties in many situations. Let us think for example about a chain of spin-1/2 particles all prepared
in their state $\ket{\downarrow}$ and able to interact by the XY interaction. 
If one now flips the spin of one of them to $\ket{\uparrow}$, this
spin excitation will propagate along the chain under the influence of the exchange interaction, 
in the same way that a single particle can tunnel between neighboring sites in a lattice. 
This transport of excitations driven by the resonant 
dipole interaction is  for example the process that takes place in photo-synthesis, where the energy deposited by light 
in a light-harvesting cell is carried towards a reaction center~\cite{Clegg2006}. 

This example suggests using a different language to describe the transport of spin excitations
under the influence of the resonant dipole interaction. 
Let us now view the $\ket{\downarrow}$-state as the absence of particle (the ``vacuum''), and rename it 
$\ket{0}$. The state $\ket{\uparrow}$ now corresponds to the presence 
of a particle, that we call $|1\rangle$. 
In a chain, the state $|0000\ldots\rangle$ corresponds to no particle on any site, while, for example $\ket{0100\ldots}$
indicates the situation where one particle is present on the second site of the chain. 
We can also consider the situation 
where two excitations are present in the chain, e.g. by preparing the state $\ket{0110\ldots}$. Importantly, these particles 
interact very strongly: as one atom cannot carry more than one excitation, it is not possible 
to find two particles on the same lattice site. Therefore spin excitations behave as 
artificial particles with infinite on-site interactions, i.e. with a hard-core constraint. 
It turns out that the spin excitations have the same commutation relations as those of bosons.  Therefore 
the problem of an ensemble of two-level Rydberg atoms interacting 
by the resonant dipole-dipole interaction can be 
mapped either onto a spin-$1/2$ XY model, or equivalently onto a system of hard-core bosons~\cite{FootnoteWignerJordan}. 
But one should keep in mind that the resonant dipole interaction that drives the transport of 
an excitation leads to a {\it single-particle} problem when considering  a single excitation. 
What makes the excitations interact is the fact that 
the atoms only have two-level, and not  the fact that the atoms carrying the excitations interact.

The first experiments performed with Rydberg atoms in a laser-cooled gas were actually a study of the 
transport of excitations driven by the dipole-dipole interaction in a situation 
where the atoms' positions are frozen~\cite{Anderson1998,Mourachko1998}. 
Since then, several experiments refined our understanding of the dynamics of 
the propagation in atomic ensembles 
with random positions~\cite{Gunter2013,Maxwell2013}. 
The case where the atoms are placed in regular arrays with individual 
control has been much less studied experimentally, and,  so far, only two experiments,
 performed in our group, have used 
the resonant dipole-dipole interaction to perform a quantum simulation. 

The first one was a proof-of-principle experiment 
demonstrating the propagation of a spin excitation in a chain of  3 atoms~\cite{Barredo2015}. 
Preparing three atoms all in the same $\ket{nD}$ Rydberg state, we flipped the state of the first one into a state
$\ket{n'P}$ and observed how the ``$P$'' excitation moved in the chain. The results, shown in Fig.~\ref{fig:fig4}(a),
indicate that despite the simplicity of the situation, the dynamics is already  non trivial: 
this comes from the fact that the two energy scales in the problem, $C_3/a^3$ and $C_3/(8a^3)$,  
lead to three incommensurate eigenvalues of the Hamiltonian.

The goal of the second experiment was to implement the Su, Schrieffer and Heeger (SSH) model, initially developed in the 
late 1970's to explain the conductivity of some organic polymers~\cite{SSH1979,Heeger1988}. In its simplest setting, 
it consists of a one-dimensional chain of sites that are coupled by an alternation of strong and weak links
and where an excitation can hop in the chain. Since then, the SSH model 
 has been recognized as one of the simplest examples of a system exhibiting topological properties. 
 Let us consider the two configurations of a finite chain  represented in Fig.~\ref{fig:fig4}b: 
 either the chain ends up with the strongest link  $J$, or with the weakest one $J'$. 
 One can show that in the first configuration, the single-particle
spectrum consists of two bands with width $J'$, separated by an energy gap $|J-J'|$. On the contrary, in the second configuration, 
two states at zero energy appear in the middle of the gap, and correspond to states localized on the edge of the chain. 
The fact that they have zero energy is rather intuitive in the extreme case where $J'=0$, as adding a particle on each 
edge does not cost energy. It turns out that this remains true even when $J'\neq0$. The two configurations correspond
to two different topological phases of the system:  it is impossible to vary the ratio $J/J'$ and continuously transform one configuration
into the other without closing the gap \cite{Asboth}. Our  group  has implemented this model in a chain of
Rydberg atoms, allowing us to study both single-particle properties, but also the genuinely 
many-body properties arising from the hard-core constraint~\cite{deLeseleuc2019}. Microwave spectroscopy was used to
measure the single-particle energy spectrum (see Fig.~\ref{fig:fig4}). We have entered the many-body regime
by preparing the ground state of the chain comprising $N/2$ excitations (where $N$ is the number of sites), using an adiabatic
preparation in the spirit of the one we used to prepare the antiferromagnetic correlations in the Ising model. We have 
characterized this many-body ground state in the topological configuration, and found it displayed a characteristic robustness with the 
respect to the breaking of certain symmetries of the Hamiltonian. The prepared state is probably the first
experimental realization of a type of topological order for bosons introduced in 2012 \cite{Wen}, 
called a symmetry-protected topological phase, which is the only one that exists in one dimension. 
  
\section{Perspectives}	

Finally, we discuss short and longer-term perspectives on the use of Rydberg atoms for quantum simulation. 
The field is rapidly evolving, and here we merely identify a few emerging directions of research. 

An obvious trend in the short term, on the technical side, is (i) to improve the fidelity of the simulations, and 
(ii) to scale up the number of atoms in the arrays. 
The first objective requires not only, at the single-particle level, 
to understand the limitations of the Rydberg excitation schemes~\cite{Leseleuc2018PRA} 
and to overcome them, for instance using different schemes for the two-photon transitions \cite{Levine2018}, 
but also, at the two-atom level, to optimize the mapping of the complex level 
structure of interacting Rydberg atoms onto simple two-level systems~\cite{Leseleuc2018PRL}. 
In addition, the possibility to scale up the number of atoms to several hundreds is one 
the crucial assets of Rydberg arrays when compared to other platforms. 
To do so, the recently demonstrated use of gray-molasses loading of optical tweezers 
\cite{Brown2019} opens up exciting prospects as, for a given trap depth, 
the required optical power per trap is strongly reduced, and, at the same time, 
the loading probability $p$ is significantly enhanced. Cryogenic platforms such as 
the ones recently developed for trapped ion chains \cite{Pagano2019} could help increase 
the atomic trapping lifetime and thus help in scaling up the atom number.  

The second short-term prospect is the extension of the techniques that have 
so far been applied to alkali atoms to new atomic species with two valence electrons. 
Arrays of single Strontium~\cite{Norcia2018,Cooper2018,MattJones} and Ytterbium~\cite{Saskin2018} 
atoms have been reported recently. Although so far no quantum simulation has been performed 
with these novel systems, the richer internal structure of these species might allow new ways 
to manipulate, control and probe them \cite{Mukherjee2011,Dunning2016}.

A longer-term goal would consist in using arrays of \textit{circular} Rydberg states 
for quantum simulation. Not only would they allow to implement more complex spin models, 
such as the the Heisenberg Hamiltonian, in a more natural way, but they may also, 
with their much longer lifetimes, open up the possibility to study long-time dynamics 
with Rydberg quantum simulators~\cite{Nguyen2017}.

Finally, one exciting prospect for Rydberg array quantum simulators 
is the fact that their applications may extend to a much broader class 
of problems than that of the mere implementation of spin Hamiltonians inspired by quantum magnetism, 
as they could be used to study optimization problems by quantum annealing~\cite{Lechner2015,Glaetzle2017}. 
While for general optimization problems Rydberg arrays may have limitations 
comparable to other platforms, they could be for instance particularly adapted to 
solving a classical combinatorial problem in graph theory, namely  
finding the Maximum Independent Set of a graph~\cite{Pichler2018}.

A hybrid, closed-loop approach, combining a Rydberg quantum simulator with 
increased degree of control and a classical computer, could be used to implement 
\textit{variational quantum simulation}. There, the quantum machine is used for  
efficiently generating many-body quantum states depending on a small number of 
variational parameters and measuring the average value of non-trivial observables, 
while classical hardware is used to optimize these parameters, making it possible to find in an interative way, e.g., 
the ground state of spin Hamiltonians that cannot be realized physically with the platform at 
hand~\cite{Kokail2019}. This type of architecture blurs the distinction between 
programmable quantum simulators and noisy, intermediate scale quantum computers 
\cite{Preskill2018}, for which Rydberg atom arrays are also a promising platform~\cite{Saffman2016,Saffman-Weiss2017}.

\begin{acknowledgments}
We thank the members of our group at Institut d'Optique, as well as all our colleagues of the Rydberg community, 
and in particular M.~Lukin, M.~Saffman, G.~Biederman, C. Gross, I.~Bloch, 
for many inspiring discussions over the years. This work benefited from financial support by
the EU (FET-Flag 817482, PASQUANS),  by ``Investissements d'Avenir'' LabEx PALM (ANR-10-LABX-0039-PALM, 
projects QUANTICA and XYLOS), and by the R\'egion \^Ile-de-France in the framework of DIM SIRTEQ (project CARAQUES).
\end{acknowledgments}


\begin{thebibliography}{99}
\bibitem{FootnoteWignerJordan}
In one dimension, they could also be
considered as free fermions, using the Wigner-Jordan transformation~\cite{Giamarchi_1D}.
\end{thebibliography}
\end{document}